\documentclass[11pt]{article}

\usepackage{amsmath, amssymb}
\usepackage{enumerate}
\usepackage{booktabs}
\usepackage{setspace}
\usepackage{appendix}
\usepackage{tikz-cd}
\usepackage[vcentermath]{youngtab}

\usepackage{color}

\numberwithin{equation}{section}

\allowdisplaybreaks

\usepackage{graphicx}
\newcommand{\p}{\partial}

\newcommand{\CC}{\mathbb{C}}

\newcommand{\YY}{\mathbb{Y}}

\newcommand{\wc}{w^{{}_{\Theta(C)}}}
\newcommand{\vc}{\rho^{{}_{\Theta(C)}}}
\newcommand{\taucone}{\tau_1^{{}_{\Theta(C)}}}
\newcommand{\tauctwo}{\tau_2^{{}_{\Theta(C)}}}
\newcommand{\tauce}{\tau_{{}_{\rm E}}^{{}_{\Theta(C)}}}

\newcommand{\beq}{\begin{equation}}
\newcommand{\eeq}{\end{equation}}

\newcommand{\nn}{\nonumber}

\newtheorem{dfn}{Definition}[section]
\newtheorem{lem}[dfn]{Lemma}
\newtheorem{prp}[dfn]{Proposition}
\newtheorem{thm}[dfn]{Theorem}
\newtheorem{rmk}[dfn]{Remark}
\newtheorem{cor}[dfn]{Corollary}

\newenvironment{prf}{\noindent {\it Proof} \ }{\hfill $\Box$}
\newenvironment{prfn}[1]{\noindent {\it Proof of #1} \ }{\hfill $\Box$}

\DeclareMathOperator{\res}{\mathrm{res}}

\begin{document}

\title{On an extension of the generalized BGW tau-function}
\author{Di Yang, \quad Chunhui Zhou}
\date{}
\maketitle
\vspace{-8mm}
\begin{center}
{\it Dedicated to the memory of Boris Anatol'evich Dubrovin, with admiration}
\end{center}
\begin{abstract}
For an arbitrary solution to the Burgers--KdV hierarchy, we 
define the tau-tuple $(\tau_1,\tau_2)$ of the solution. We show that 
the product~$\tau_1\tau_2$ admits Buryak's residue formula. Therefore, according to Alexandrov's theorem, 
$\tau_1\tau_2$ is a tau-function of the KP hierarchy. 
We then derive a formula 
for the affine coordinates for the point of the Sato Grassmannian corresponding to the 
tau-function~$\tau_1\tau_2$ explicitly in terms of those 
for~$\tau_1$. 
Applications to the analogous open extension of the generalized 
BGW tau-function and to the open partition function are given.
\end{abstract}

\tableofcontents

\section{Introduction}
The Br\'ezin--Gross--Witten (BGW) partition function was introduced in~\cite{BG, GW}; see also~\cite{A3, DN, DYZ, MMS,M1}. 
Recently, Norbury~\cite{N} constructed a cohomology class (called the {\it Theta-class})
on the moduli space of closed Riemann surfaces, 
and indicated~\cite{N} that the partition function of the intersection numbers 
of the Theta-class coupling the well-known $\psi$-classes coincides with the BGW partition function. 
Recalling the construction about the 
moduli space of open Riemann surfaces~\cite{PST} as well as the 
 open extension of the partition function of the $\psi$-class intersection numbers~\cite{B2,PST}, 
we consider in this paper the analogous open extension of the BGW partition function.

To illustrate more the motivations of the paper and of the terminology, 
let us provide some geometric ideas~\cite{Du1, DZ, W}.
One important observation is due to Paolo Rossi. Rossi finds\footnote{One of the authors D.Y. is grateful to 
Paolo Rossi and Sasha Buryak
for sharing with him this interesting observation during a workshop at the University of Leeds in 2018.} 
that the {\it open} WDVV equations given by Horev and Solomon~\cite{HS} for $(n+1)$ observables 
could be viewed as the WDVV equations of an $(n+1)$-dimensional {\it generalized} 
Frobenius manifold (with the existence-of-potential condition dropped, sometimes called flat F-manifold), 
which is a one-dimensional extension of 
an $n$-dimensional {\it Frobenius manifold}. Nevertheless, this type of generalized Frobenius 
manifold has a {\it vector} potential, from which the genus zero part of the abstract 
open-and-closed descendent invariants can be re-constructed~\cite{BaBur}. For more details about 
Rossi's observation as well as its outcome see~\cite{BaBur}. Rossi's observation  
gives rise to a {\it tau-tuple} of the topological solution
to the corresponding dispersionless open hierarchy. It also naturally invokes, following Witten and 
Dubrovin--Zhang~\cite{BaBur,Du1,DLYZ,DZ, W}, the tau-tuple of the topological solution to 
the {\it topological deformation} of the dispersionless open hierarchy. This is non-trivial already when $n=1$.
The structure of a tau-tuple (being referred to as the tau-structure) will capture the geometry for the open-and-closed descendent invariants, 
and enables one, due to its differential-polynomial nature, 
to deal with an arbitrary solution to the open hierarchy (not just the topological solution giving the open-and-closed descendent invariants). 
Then, if the tau-function of some other solution to the integrable hierarchy of topological type 
of a Frobenius manifold has a topological expansion possessing enumerative meanings 
on the moduli space of closed Riemann surfaces then its analogous open 
extension could be expected to have a meaning for the moduli space of open-and-closed Riemann surfaces. 
In this paper we will define the tau-structure for the Burgers--KdV hierarchy~\cite{B1}, and use it 
to give the analogous open extension for the BGW partition function. (The latter is proved to be a particular 
tau-function for the KdV hierarchy~\cite{A3, DN, N} and possess enumerative meanings 
on the moduli space of closed Riemann surfaces.) Noting that there is a one-parameter generalization of the BGW partition 
function given in~\cite{A3}, we will actually study in detail 
the analogous open extension for an arbitrary value of the parameter, although 
we do not know if the generalized BGW partition function with the general parameter 
is connected to the moduli space of curves.
We will also derive some general explicit formulae for the open extensions of an arbitrary solution 
using Buryak's residue formula~\cite{B2}. 
According to our knowledge, our 
explicit simplification is new even for the open partition function~\cite{B2}.

Denote by $\mathcal A_{(w,\rho)}$ the ring of polynomials of $w_x,\rho_x,w_{xx},\rho_{xx}, \dots$ with coefficients being smooth functions of $w,\rho$.   
The Burgers--KdV integrable hierarchy~\cite{B1} is defined as the following pairwise commuting system of non-linear PDEs:
\begin{align}
&\frac{\p w}{\p q_n}=\p_x K_{n}, \label{eq-kdv}\\
&\frac{\p \rho}{\p q_n}=\p_x R_{n}, \label{eq-bur}
\end{align}
where $n\geq1$, and $K_n, R_n \in \mathcal{A}_{(w,\rho)}$ are defined recursively by
\begin{align}
&K_1=w,\quad K_2=0,\quad R_1=\rho, \quad R_2=\rho^2+\rho_x+2w,\\
&\p_x K_{2j+1}=\left(2w \p_x+ w_x+\frac14\p_x^3\right) K_{2j-1}, \quad K_{2j}=0,\\
&R_{2j+1}=\left( \p_x^2+2\rho \p_x+\rho^2+\rho_x+2w\right)R_{2j -1}+\left(\rho+\frac32 \p_x\right)K_{2j - 1},\\
&R_{2j+2}=\left( \p_x^2 +2 \rho \p_x + \rho^2+\rho_x + 2w \right) R_{2j}.
\end{align}
Here, $j\geq 1$, and the integration constants in the recursive procedure 
of solving $K_{2j+1}$ are chosen to be zero. The $n=1$ flow reads 
\[
\frac{\p w}{\p q_1}=w_x, \quad  
\frac{\p \rho}{\p q_1}=\rho_x.
 \] 
Therefore, we can identify the variable~$q_1$ with~$x$.
Note that equations~\eqref{eq-kdv} are the KdV integrable hierarchy. 
Hence, the Burgers--KdV hierarchy can be viewed as an
integrable extension of~\eqref{eq-kdv}; 
for more details about the Burgers--KdV hierarchy see~\cite{B1}.

Recall that the canonical tau-structure~\cite{DYZ,DZ}
of the KdV hierarchy is defined as the unique family of elements $(\Omega_{i,j})_{i,j\geq1}\in \mathcal{A}_{w}$ 
satisfying
\beq\label{taustruc}
\Omega_{1,1}=w,\quad \Omega_{i,j}=\Omega_{j,i},\quad 
\frac{\p \Omega_{i,j}}{\p q_{2k-1}}=\frac{\p \Omega_{i,k}}{\p q_{2j-1}},\qquad \forall\,i,j,k\geq1
\eeq
along with the normalization condition $\Omega_{i,j}|_{w_{kx}=0,k\geq 0}=0$.  
Here, $\mathcal{A}_w\subset \mathcal{A}_{(w,\rho)}$ denotes the ring of polynomials of $w_x,w_{xx},\dots$ with coefficients being smooth functions in~$w$.  For the existence of $\Omega_{i,j}$ see for example~\cite{DYZ,DZ}. 
Inspired by the open and closed intersection numbers~\cite{BeY,B1,B2,PST} 
and by the tau-structure of the Burgers hierarchy~\cite{DY}, 
we will define the tau-tuple for the Burgers--KdV hierarchy.  
It is shown in~\cite{B1} that
\beq
\frac{\p R_i}{\p q_j}=\frac{\p R_j}{\p q_i},\qquad \forall\,i,j\geq1.
\eeq
By using these relations as well as the relations~\eqref{taustruc} we arrive at the following lemma.

\begin{lem}\label{lem-deftau}
For an arbitrary solution $(w,\rho)$ in $\CC[[{\bf q}]]^2$ to the Burgers--KdV hierarchy, 
there exist $\tau_1,\tau_2\in\CC[[{\bf q}]]^2$, such that 
\begin{align}
&\frac{\p^2 \log\tau_1}{\p q_{2i-1} \p q_{2j-1}}=\Omega_{i,j}, \quad
\frac{\p^2 \log\tau_1}{\p q_{2i} \p q_{j}}=0,\quad
\forall \, i,j\geq1,\label{eq-tau1}\\
&\frac{\p \log \tau_2}{\p q_{i}}=R_{i},\quad \forall \, i\geq1.\label{eq-tau2}
\end{align}
Moreover, the tuple $(\tau_1,\tau_2)$ is uniquely determined by $(w,\rho)$ up to the freedom described by 
\beq\label{eq-cons}
\left(\tau_1,\tau_2\right) \mapsto \left( \tau_1 e^{\beta_0+\sum_{n\geq1} \beta_n q_n},\, \tau_2 e^{\alpha_0}\right),  
\eeq
where $\alpha_0$ and $\beta_n$, $n\geq 0$ are arbitrary constants. 
\end{lem}
\noindent We call $(\tau_1,\tau_2)$ the \emph{tau-tuple} of the solution $(w,\rho)$ to the Burgers--KdV hierarchy, 
and call the product $\tau_1 \tau_2=:\tau_{{}_{\rm E}}$ the {\it tau-function} for the Burgers--KdV hierarchy. 
Equations~\eqref{eq-tau1}--\eqref{eq-tau2} for $i=j=1$ read 
\beq\label{definitiontau12}
w=\p_x^2 \log\tau_1,\quad \rho=\p_x \log \tau_2.
\eeq

As it is proved in~\cite{BDY}, the definition of a Dubrovin--Zhang type tau-function~$\tau_1$ for the KdV hierarchy 
is equivalent to that of a Sato type tau-function; see also~\cite{BBT, D}.
We proceed to the study of the tau-tuple from the points of view of the Sato theory. 
Let us first recall some notations. 
Denote by $\YY$ the set of all the partitions. 
For each partition $\lambda\in\YY$, denote by $\ell(\lambda)$ the length of $\lambda$, by $|\lambda|$ the weight, 
and by $(m_1,\dots,m_r|n_1,\dots,n_r)$ the Frobenius notation (cf.~\cite{M}).
The Schur polynomial associated to $\lambda$ is defined as follows:
\beq\label{eq-sch-1}
s_\lambda:=\det (h_{\lambda_i-i+j})_{1\leq i,j \leq \ell(\lambda)},
\eeq
where $\{h_k\}_{k\geq0}$ are polynomials in $\CC[{\bf q}]$ defined by the generating function
\beq\label{eq-h}
\sum_{k\geq0}h_k z^k=\exp\Big(\sum_{n\geq1} q_n z^n\Big).
\eeq
According to the Sato theory, the component $\tau_1$, 
can be expressed as the linear combination of the 
Schur polynomials with 
coefficients being the Pl\"ucker coordinates, i.e.,
\beq\label{eq-sch-exp}
\tau_1= c_0 \sum_{\lambda\in\YY} \pi_\lambda s_\lambda.
\eeq
Here, $c_0$ is a nonzero constant often taken as~1, and 
\beq \label{ZhouBaloghYang}
\pi_\lambda := (-1)^{\sum_{i=1}^r n_i}\det(A_{m_i,n_j})_{1\leq i,j\leq r}
\eeq
with 
$A_{m,n}$, $m,n\geq0$ being the {\it affine coordinates}~\cite{BY,EH,Z1,Z2} for the point of Sato Grassmannian corresponding to~$\tau_1$. 
In this paper, we will derive explicit expressions for~$\tau_1 \tau_2$. 

Before presenting the results, we introduce further some notations. 
Note that the Schur polynomials form a basis of $\CC[[{\bf q}]]$, and define a sequence of linear operators 
$T_n:\CC[[{\bf q}]]\rightarrow \CC[[{\bf q}]]$ 
in the following way:
\beq\label{deftn}
T_n(s_\lambda)=\left\{
\begin{array}{cc}
(-1)^i s_{(\lambda_1-1,\dots,\lambda_i-1,n+i,\lambda_{i+1},\dots,\lambda_{\ell(\lambda)})}, 
\quad 
&\text{if } \lambda_i>n+i\geq\lambda_{i+1},\,\text{for some $i$},\\
0,\quad &\text{otherwise}.
\end{array}
\right.
\eeq
Here we set $\lambda_0=+\infty$ and $\lambda_{\ell(\lambda)+1}=0$. 
The operator $T_0$ has the alternative expression:
\begin{align}
& T_0 \left( s_{(m_1,\dots,m_r |n_1,\dots,n_r)}\right)
=\left\{
\begin{array}{cc}
(-1)^{r} s_{(m_1-1,\dots,m_r-1|n_1+1,\dots,n_r+1)},\quad & m_r\geq1,\\
0,\quad &m_r=0.
\end{array}
\right.\label{eq-ts}
\end{align}
Note that for each $n\geq0$, 
the linear operator $T_n$ induces the linear transformation on  
${\rm Span}_{\CC}(\YY)$, denoted again by $T_n$. 
The first result of the paper is given by the following proposition. 

\begin{prp}\label{prp-qtq}
Let $(w,\rho)$ be an arbitrary solution in $\CC[[{\bf q}]]^2$ to the Burgers--KdV hierarchy~\eqref{eq-kdv}--\eqref{eq-bur}, 
and $(\tau_1,\tau_2)\in\CC[[{\bf q}]]^2$ its tau-tuple
with~$\tau_2({\bf 0})$ chosen as~$1$.
There exists a unique sequence of numbers $b_1,b_2,b_3,\dots$, 
such that
\beq\label{eq-tau12}
\tau_1 \, \tau_2=Q\circ  T_0 \circ Q^{-1} (\tau_1),
\eeq
where $Q=e^{\sum_{k\geq1}k b_k q_k}$ denotes the multiplication operator on $\CC[[{\bf q}]]$.
\end{prp}
The proof is in Section~\ref{sec3}. 

Noticing that $Q\circ  T_0 \circ Q^{-1} $ is a linear operator and using~\eqref{eq-sch-exp}, 
we will simplify~\eqref{eq-tau12} via computing the action of $Q\circ  T_0 \circ Q^{-1}$ on each single Schur polynomial. 
Let us present the results here, and leave the details of calculations to Section~\ref{sec3}.
For $\lambda\in\YY$, define $\pi_{-\lambda}:=-\pi_\lambda$. 
Let $c_{\lambda\mu}^\nu$, $\lambda,\mu,\nu\in\YY$ denote the Littlewood--Richardson coefficients~\cite{LR}:
\beq
s_\lambda \, s_\mu=:\sum_{\nu\in\YY} c_{\lambda\mu}^\nu s_\nu.
\eeq
For $\lambda,\mu,\nu\in\YY$, we set $c_{\lambda,-\mu}^\nu:=-c_{\lambda\mu}^\nu$. 
Define a sequence of numbers $g_0,g_1,g_2,g_3,\dots$ by
\beq\label{bandg}
\sum_{n\geq0} \,g_n \, z^{-n} \,:=\, e^{\sum_{k\geq1}b_k z^{-k}}.
\eeq 

\begin{prp}\label{prp-sch-e}
The tau-function $\tau_{{}_{\rm E}}$ has the expression
\beq\label{tauea}
\tau_{{}_{\rm E}} =
\sum_{\lambda\in\YY}  s_\lambda  
\sum_{\substack{0\leq k\leq |\lambda|\\ \mu \in \mathcal{T}_k^{-1}(\lambda)}} \, g_k \, \pi_{\mu},
\eeq
where $\mathcal T_k:\YY\cup(-\YY)\rightarrow\YY\cup(-\YY)$ is defined as $T_k|_{\YY\cup(-\YY)}$.
Alternatively, denoting ${\bf b}=(b_1,b_2,b_3,\dots)$, we have
\beq\label{tauea-1}
\tau_{{}_{\rm E}}=
\sum_{\lambda\in\YY}  s_\lambda
\sum_{\substack{\alpha,\mu,\nu\in\YY \\ \beta\in\YY\backslash{{\rm Ker}}(T_0)}}
\pi_\alpha \, c_{\mu\alpha}^\beta \, c_{\nu,T_0(\beta)}^\lambda \, 
s_ \mu(-{\bf b}) \, s_\nu ({\bf b}).
\eeq
 \end{prp}
 
For a hook partition~$\lambda$ with the Frobenius notation~$(m|n)$, we have
\beq
 \mathcal{T}_k^{-1}(\lambda)=\left\{
 \begin{array}{ccc}
 & \{(0)\}, \quad & n=0,\,k=m+1,\\
 & \{(0|n-1)\}, \quad & n\geq1, k=m+1,\\
 & \{-(m+1|n-1)\},\quad & n\geq1, k=0, \\
 &\varnothing, \quad & \text{otherwise}.
 \end{array}
 \right.
\eeq
Here, $(0)$ denotes the partition of~0.
Then from~\eqref{tauea} of Proposition~\ref{prp-sch-e} we see that 
 the coefficient of~$s_{(m|n)}$ in~$\tau_{{}_{\rm E}}$, denoted by~$c_{m,n}$, has 
 the more explicit expression:
\beq\label{primea}
 c_{m,n}=
 \left\{
 \begin{array}{cc}
g_{m+1} ,\quad  &n=0,\\
(-1)^n \left(g_{m+1}A_{0,n-1}-A_{m+1,n-1}\right),\quad &n\geq1.
\end{array}
\right.
\eeq
Combining with Theorem~\ref{thm-kp-tau} and 
 the formula similar to~\eqref{eq-sch-exp} for the KP hierarchy (cf.~\cite{BY, DJKM, D, EH, IZ, S, Z2}), we arrive at 
 the following corollary.
 \begin{cor}\label{coraffine}
 The tau-function $\tau_{{}_{\rm E}}$ has the following explicit expression: 
 \beq\label{tauea-2}
 \tau_{{}_{\rm E}} \,=\, \sum_{\lambda\in\YY} \, \pi'_\lambda \, s_\lambda.
 \eeq
 Here for a partition $\lambda=(m_1,\dots,m_r|n_1,\dots,n_r)$,
  $\pi'_\lambda=\det (c_{m_i,n_j})_{1\leq i,j\leq r}$ with $c_{m,n}$ defined by~\eqref{primea}.
 \end{cor}

Both Proposition~\ref{prp-sch-e} and Corollary~\ref{coraffine} can be used for efficient computations of~$\tau_{{}_{\rm E}}$.

	\begin{rmk}
		Propositions~\ref{prp-qtq}, \ref{prp-sch-e} and Corollary~\ref{coraffine}
		could be generalized to the KP hierarchy.
		In fact, let~$\tau_1$ be an arbitrary tau-function of the KP hierarchy, and let $\tau_{{}_{\rm E}}$
		be defined via Buryak's residue formula:
		\[
		\tau_{{}_{\rm E}}:=-\res_{z=\infty}
		\biggl(\sum_{n\geq0}g_n z^{-n}\biggr) \, 
		\tau_1\Bigl(q_1-\frac1z,q_2-\frac1{2z^2},\dots\Bigr) \, 
		e^{\sum_{k\geq1}q_k z^k} \, \frac{dz}{z},
		\]
		where $g_0=1$ and $g_2,g_3,\dots$ are arbitrary constants.
		Then $\tau_{{}_{\rm E}}$ is also a tau-function of the KP hierarchy~\cite{A2} 
		(see Theorem~\ref{thm-kp-tau} of Appendix~\ref{app} for the details and a new proof), 
		and as a generalization of Proposition~\ref{prp-qtq} this tau-function
		can be written into
		\[
		\tau_{{}_{\rm E}}=Q \circ T_0 \circ Q^{-1}(\tau_1),
		\]
		where the coefficients in
		$Q=e^{\sum_{k\geq1}k b_k q_k}$
		are determined by the relation \[e^{\sum_{k\geq1}b_k z^{-k}} \,=\, \sum_{n\geq0} \, g_n \, z^{-n}.\]
		Moreover, formulae~\eqref{tauea}, \eqref{tauea-1} and~\eqref{tauea-2} (with $c_{m,n}$ defined by~\eqref{primea})
		in Proposition~\ref{prp-sch-e} and Corollary~\ref{coraffine}
		are still valid. 
		We will give some combinatorial identities 
		as applications of these results at the end of Appendix~\ref{app}.
	\end{rmk}

We proceed with defining the extension of the generalized BGW tau-function. 
Consider the solution to the Burgers--KdV hierarchy~\eqref{eq-kdv}--\eqref{eq-bur} specified by the initial condition
\beq\label{eq-ini}
\left. w\right|_{q_2=q_3=\cdots=0}=\frac{C}{(1-x)^2},\quad 
\left. \rho\right|_{q_2=q_3=\cdots=0}=0,
\eeq
where $C$ is an arbitrarily given constant ($N=\sqrt{\frac14-2C}$ corresponds to the parameter in~\cite{A3}).
We call this solution the generalized 
BGW solution to the Burgers--KdV hierarchy, 
denoted by
$(\wc,\vc)$.
Let~$(\taucone,\tauctwo)$ be the tau-tuple of~$(\wc,\vc)$. 
As in~\cite{A3, BR, DN, DYZ} we normalize the first component~$\taucone$ by
\beq\label{eqstr1}
\sum_{k\geq 0} \, (2k+1) \, (q_{2k+1}-\delta_{k,0}) \, \frac{\p \taucone}{\p q_{2k+1}} \,+\, C\, \taucone \,=\, 0
\eeq
and by requiring $\p \taucone/\p q_{2n}=0$, call it the {\it generalized BGW tau-function}. (Note that $\taucone$ is 
uniquely determined up to a constant factor that is irrelevant of the study.)
We normalize the second component~$\tauctwo$ by requiring 
\beq\label{conditiontau20}
\tauctwo({\bf 0}) = 1.
\eeq
We note that when the parameter $C$ is taken to be $1/8$, 
the first component $\taucone$ becomes the BGW tau-function. 
Norbury~\cite{N} proves the following identity:
\beq
\tau_1^{{}_{\Theta(\frac18)}}
= \exp\left(\sum_{n\geq0}\sum_{i_1,\dots,i_n\geq0}\frac{1}{n!} \,
\prod_{k=1}^n(2i_k+1)!! \, q_{2i_k+1}\int_{\overline{\mathcal M}_{g,n}}\Theta_{g,n}\psi_1^{i_1}\cdots\psi_n^{i_n}\right).
\eeq
Here, $\psi_i$ denotes the first Chern class of the $i$th tautological line bundle on~$\overline{\mathcal{M}}_{g,n}$, 
and $\Theta_{g,n}$ denotes the Theta-class~\cite{N}.
Our next result is given by the following theorem. 
\begin{thm}\label{thm-main}
Define a sequence of numbers $g_n^{{}_{\Theta(C)}}$ by 
\beq\label{eq-defgn}
g_0^{{}_{\Theta(C)}}=1,\quad g_n^{{}_{\Theta(C)}} \,= \,(n-1)! \, \sum_{k=1}^n \frac{1}{k!(k-1)!}\prod_{i=1}^k\left(C+\frac{i(i-1)}{2}\right),\quad n \geq1.
\eeq
Define $b_1,b_2,b_3,\dots$ via $\sum_{k\geq1}b_k z^{-k}=\log \left(\sum_{n\geq0} g_n^{{}_{\Theta(C)}} z^{-n}\right)$, and denote by $Q=e^{\sum_{k\geq1}k b_k q_k} $ the multiplication operator. 
The following identity holds true:
\beq\label{eq-bgw12}
\tauctwo \,=\, \frac1{\taucone} \, Q\circ T_0 \circ Q^{-1}(\taucone).
\eeq 
Moreover,  $\tauce:=\taucone\tauctwo$
satisfies the following linear equation:
\beq\label{eqstr2}
\sum_{n\geq1} \, n \, \tilde q_n \, \frac{\p\tauce}{\p q_n} \,+\, C \, \tauce \;=\; 0,\quad \tilde q_n=q_n-\delta_{n,1}. 
\eeq
\end{thm}

Recall that based on~\cite{BY, Z2} the explicit expression for the affine coordinates~$A_{m,n}^{{}_{\Theta(C)}}$ 
of the point of the Sato Grassmannian corresponding to~$\taucone$ was  
obtained in~\cite{DYZ} (cf.~also~\cite{Fu}):
\beq\label{bgwa-2}
A_{m,n}^{{}_{\Theta(C)}}=
(-1)^m\sum_{\substack{r\geq m+1,s\geq0\\r+s=m+n+1}}\frac{r-s}{r+s} \, a_r a_s, \quad
a_k:=\frac{(-1)^k}{k!}\prod_{i=1}^k\left(C+\binom i 2\right).
\eeq
Define $\pi^{{}_{\Theta(C)}}_{\lambda}$ via~\eqref{ZhouBaloghYang} with $A_{m,n}$ replaced by~$A_{m,n}^{{}_{\Theta(C)}}$. 
Combining with~\eqref{tauea} 
we arrive at
\begin{cor}\label{corbgw}
The tau-function $\tauce$ has the form
\beq\label{bgw-sch}
\tauce
=\sum_{\lambda\in\YY}  s_\lambda 
\sum_{\substack{0\leq k\leq |\lambda|\\ \mu \in \mathcal T_k^{-1}(\lambda)}} g_k^{{}_{\Theta(C)}} \pi^{{}_{\Theta(C)}}_\mu.
\eeq
\end{cor}
For the reader's convenience we list the first few coefficients of $\tauce$ in Appendix~\ref{app-a}.

Propositions~\ref{prp-qtq} and~\ref{prp-sch-e} also apply to the topological solution to the Burgers--KdV hierarchy, 
denoted by $(w^{{}_{\rm WK}},\rho^{{}_{\rm WK}})$, 
which is specified by the initial condition
\beq\label{topo}
\left. w^{{}_{\rm WK}} \right|_{q_2=q_3=\cdots=0}=x,\quad
\left. \rho^{{}_{\rm WK}} \right|_{q_2=q_3=\cdots=0}=0.
\eeq
This solution governs the open and closed intersection numbers~\cite{B1,B2,PST}.
Let $\tau_1^{{}_{\rm WK}}$ be the tau-function of the solution $w^{{}_{\rm WK}}$ to the KdV hierarchy, normalized by 
the string equation:
\beq\label{string}
\sum_{n\geq0} \, (2n+1)\left( q_{2n+1}-\delta_{n,1}\right)\frac{\p \tau_1^{{}_{\rm WK}}}{\p q_{2n+1}} \,+\, \frac{x^2}{2} \tau_1^{{}_{\rm WK}} \,=\, 0.
\eeq
The explicit expression for the affine coordinates~$A_{m,n}^{{}_{\rm WK}}$ 
for the point of the Sato Grassmannian corresponding to~$\tau_1^{{}_{\rm WK}}$ was  
first obtained in~\cite{Z1} and later re-proved in~\cite{BY}:
\begin{align}
 &A_{3m-1,3n}^{{}_{\rm WK}}=A_{3m-3,3n+2}^{{}_{\rm WK}} \nn \\
=&\frac{(-1)^n}{36^{m+n}}\frac{(6m+1)!!}{(2m+2n)!}
\prod_{j=0}^{n-1}(m+j)\prod_{j=1}^n(2m+2j-1) \left(B(n,m)+\frac{2^n (6n+1)!!}{(6m+1)(2n)!}\right), \label{wka-2}\\
 &A_{3m-2,3n+1}^{{}_{\rm WK}} \nn \\
=&\frac{(-1)^{n+1}}{36^{m+n}}\frac{(6m+1)!!}{(2m+2n)!}
 \prod_{j=0}^{n-1}(m+j)\prod_{j=1}^n(2m+2j-1) 
 \left(B(n,m)+\frac{2^n (6n+1)!!}{(6m-1)(2n)!}\right), \label{wka-3}
\end{align}
where $n\geq0$, $m\geq1$ and $B(n,m)$ are defined as
\beq
B(n,m):=\frac16\sum_{j=1}^n 108^j \frac{2^{n-j}(6n-6j+1)!!}{(2n-2j)!}\frac{\Gamma(m+n+1)}{\Gamma(m+n+2-j)}.
\eeq
Recall from~\cite{B2} that the values $g_n$ for the topological solution are given by 
\beq\label{wka-4}
g_n^{{}_{\rm WK}}=\left\{ \begin{array}{cl} 
\sum_{i=0}^m \frac{3^i (6m-6i)!}{288^{m-i}(2m-2i)!(3m-3i)!} \prod_{j=1}^i (m+\frac12-j), & n=3m, \\ 
0, & {\rm otherwise}.\\
\end{array} \right.
\eeq
Define $\pi_\lambda^{{}_{\rm WK}}$ via~\eqref{ZhouBaloghYang} with $A_{m,n}$ replaced by~$A_{m,n}^{{}_{\rm WK}}$. 
Using~\eqref{tauea} 
we then arrive at 
\begin{cor}\label{corwk}
The power series $\tau_{{}_{\rm E}}^{{}_{\rm WK}}$ has the following expression:
\beq
\tau_{{}_{\rm E}}^{{}_{\rm WK}}
=
\sum_{\lambda\in\YY}  s_\lambda 
\sum_{\substack{0\leq k\leq |\lambda|\\ \mu\in\mathcal T_k^{-1}(\lambda)}} g_k^{{}_{\rm WK}} \pi^{{}_{\rm WK}}_\mu.
\eeq
\end{cor}

\noindent{\bf Organization of the paper} 
In Section~\ref{sec2}, 
we will recall about some basics about the KdV hierarchy.
In Section~\ref{sec3}, 
we prove Propositions~\ref{prp-qtq} and~\ref{prp-sch-e}. 
In Section~\ref{sec4}, 
we prove Theorem~\ref{thm-main}. 
Besides, we give a new proof of a result of Alexandrov in Appendix~\ref{app}.

\medskip

\noindent{\bf Acknowledgement} 
We are grateful to Si-Qi Liu and Youjin Zhang for very helpful suggestions and encouragements. 
D.Y. is grateful to Boris Dubrovin and Don Zagier for their advice. 
We would like to thank the anonymous referees for useful comments that help to improve the presentation of the paper. 
The work receives funding from the 
National Key Research and Development Project ``Analysis and Geometry on Bundles" 
 SQ2020YFA070080. D.Y. is supported in part by another national research grant.

\section{A brief review of the KdV hierarchy}\label{sec2}
In this section, we recall some basic facts about the KdV hierarchy~\eqref{eq-kdv}. 
It is well known that this hierarchy can be written into the Lax form
\beq\label{eq-kdvlax}
\frac{\p  L}{\p q_n}=\left[\left( L^{\frac {n} 2}\right)_+,\, L\right],\quad n\geq1,
\eeq
where $ L:=\p_x^2+2 w(x)$. Here, for the meaning of $L^{n/2}$ see e.g.~\cite{D}.
For any solution $w\in\CC[[{\bf q}]]$ to the KdV hierarchy\footnote{Usually the even flows are not included. 
Since these flows vanish and are therefore trivial, we have added them in~\eqref{eq-kdv}.
In other words, we understand in the way that~\eqref{eq-kdv} provide solutions for the KP hierarchy.
}, 
there exists~\cite{D,DYZ} a pseudo-differential operator 
$P=\sum_{i\geq0}\phi_i \p_x^{-i}\in\CC[[{\bf q}]][[\p_x^{-1}]]$, $\phi_0:=1$ 
satisfying the followings:
\begin{align}
&  L=P\circ \p_x^2 \circ P^{-1}, \label{p1} \\
& \frac{\p P}{\p q_n}=-\big( L^{\frac n2}\big)_- \circ P,\quad \forall \, n\geq1. \label{eq-pevo}
\end{align}

The formal adjoint operator of $P$ is defined by 
\beq
P^*:=\sum_{i\geq0}(-\p_x)^{-i}\circ \phi_i=:\sum_{n\geq 1}\phi_i^* \p_x^{-i}.
\eeq 
For each solution $w$, define the wave function $\psi$ and its dual by
\begin{align}
&\psi:=\psi({\bf q};z)=\sum_{i\geq0}\phi_i z^{-i} e^{\sum_{n\geq1}q_n z^n}, \label{defpsi}\\
&\psi^*:=\psi^*({\bf q};z)=\sum_{i\geq0}\phi_i^* z^{-i} e^{-\sum_{n\geq1}q_n z^n}. \label{defpsistar}
\end{align}
Then $\psi$ and $\psi^*$ satisfy the following linear equations:
\begin{align}
& L \psi = z^2 \psi,\quad  L \psi^*=z^2 \psi^*, \label{eq-lpsi}\\
&\frac{\p \psi}{\p q_n}=\left( L^{\frac n 2}\right)_+ \psi,\quad 
 \frac{\p \psi^*}{\p q_n}=\left(( L^*)^{\frac n 2}\right)_+ \psi^*,\quad \forall\,n\geq1. 
\label{eq-psi}
\end{align}
The following lemma was proven in e.g.~\cite{DJKM}.
\begin{lem}[\cite{DJKM}]\label{lem-bieq}
The following bilinear identity holds true:
\beq\label{eq-bili}
\res_{z=\infty}\psi({\bf q} ;z) \, \psi^*({\bf q}';z) dz=0, \quad \forall\, {\bf q}=(q_1,q_2,\dots), {\bf q}'=(q_1',q_2',\dots ).
\eeq
\end{lem}

According to the Sato theory~\cite{DJKM,D,S}, there exists a (unique up to a multiplicative constant) formal series 
$\tau({\bf q})$ satisfying 
\begin{align}
&\psi\left({\bf q};z\right)=\frac{\tau({\bf q}-[z^{-1}])}{\tau({\bf q})} e^{\sum_{n\geq1}q_n z^n},
\quad \psi^*\left({\bf q};z\right)=\frac{\tau({\bf q}+[z^{-1}])}{\tau({\bf q})} e^{-\sum_{n\geq1}q_nz^n},\label{eq-psitau}
\end{align}
where $[z^{-1}]:=(\frac1z,\frac1{2z^2},\frac1{3z^3},\dots)$.
The series $\tau({\bf q})$ is called the {\it Sato type tau-function} of the solution~$w$ 
for the KdV hierarchy, which will play an important role in the later sections.
Observe that the above $\phi_i$ and $\phi_i^*$, $i\geq0$ 
do not depend on $q_2,q_4,q_6,\dots$. 
Therefore, if we view $q_2,q_4,q_6,\dots$ as constants, 
then as it was proved in~\cite{DYZ} the $\psi$ and~$\psi^*$ defined by~\eqref{defpsi}--\eqref{defpsistar} 
 satisfy the pair condition of~\cite{DYZ}. This statement is applied in several places of this paper.

\section{Proofs of Propositions~\ref{prp-qtq} and~\ref{prp-sch-e}}\label{sec3}
The goal of this section is to prove Propositions~\ref{prp-qtq} and~\ref{prp-sch-e}.

The following lemma, that will play a fundamental role, generalizes a result of Buryak~\cite{B2}.
\begin{lem}\label{lem-gz}
Let $(w,\rho)\in\CC[[{\bf q}]]^2$ be an arbitrary solution to the Burgers--KdV hierarchy, and 
let $(\tau_1,\tau_2)\in\CC[[{\bf q}]]^2$ be its tau-tuple
with a fixed choice~\eqref{eq-cons}. 
There exists a unique series
\beq\label{gzform}
g(z) \,=\, \sum_{n\geq0} \, g_n \, z^{-n}\in\CC[[z^{-1}]],\quad g_n\in\CC, \,g_0\neq 0,
\eeq
such that the component $\tau_2$ has the expression
\beq\label{eq-tau-gz}
\tau_2=-\res_{z=\infty} g(z) \, \frac{\tau_1\left({\bf q}-[z^{-1}]\right)}{\tau_1({\bf q})} \, e^{\sum_{n\geq1}q_n z^n}\frac{dz}{z}.
\eeq
\end{lem}

\begin{prf}
According to~\cite{BDY}, 
the Dubrovin--Zhang type tau-function~$\tau_1$ of the solution~$w$ for the KdV hierarchy  
defined by~\eqref{eq-tau1} is also a Sato type tau-function of~$w$, i.e., the $\psi$ defined by 
\[
\psi({\bf q};z):= \frac{\tau_1({\bf q}-[z^{-1}])}{\tau_1({\bf q})} \, e^{\sum_{n\geq1}q_n z^n}
\]
is a wave function~\eqref{eq-psi} of the KP hierarchy. Let us define a sequence of numbers $g_n$, $n\geq 0$:
\begin{align}
&g_0=\tau_2({\bf 0}),\label{gn1}\\
&g_n=
\p_x^n|_{{\bf q}={\bf 0}} (\tau_2)
+\sum_{k=0}^n\sum_{i=0}^{{\rm min}\{k,n-1\}} 
\res_{z=\infty}\frac{g_i \, dz}{z^{k-i+1}} \, 
\p_x^{n-k}\big|_{x=0}\left(\frac{\tau_1(x-\frac1z,-\frac1{2z^2},\dots)}{\tau_1(x,0,\dots)}\right), 
\quad n\geq1.\label{gn2}
\end{align}
Define $g(z)=\sum_{n\geq0} g_n z^{-n}$, and define $\tilde\tau_2\in\CC[[{\bf q}]]$ by 
$\tilde\tau_2 = -\res_{z=\infty}
g(z)\psi({\bf q};z)\frac{dz}{z}$. From the definition of~$g_n$ we have
\beq\label{tau22}
\tilde \tau_2(x,0,\dots)=\tau_2(x,0,\dots).
\eeq 
Using~\eqref{eq-psi} we know that 
$\forall\,n\geq1$,
\beq\label{bur1}
\frac{\p \tilde \tau_2}{\p q_n}
\, = \, \bigl(L^{\frac n 2}\bigr)_+(\tilde \tau_2),
\eeq
where $L=\p_x^2 + 2w$. 
According to~\eqref{eq-tau2} and~\cite{B1}, the second component~$\tau_2$ satisfies 
\beq\label{bur2}
\frac{\p\tau_2}{\p q_n}
\,=\, \bigl(L^{\frac n 2}\bigr)_+ (\tau_2),\quad \forall\,n\geq1.
\eeq 
Combining with formulae~\eqref{tau22}--\eqref{bur2}, we obtain $\tau_2=\tilde\tau_2$. 
The uniqueness follows from the requirement~\eqref{eq-tau-gz} 
(in fact restricting to $q_2=q_3=\cdots=0$ already implies the uniqueness of~$g_n$).
The lemma is proved.
\end{prf}

The following lemma gives explicit computation on the maps $T_n$ in terms of the Schur basis.
\begin{lem}\label{lemtnm}
For each $n\geq0$ and $\lambda\in\YY$, 
we have
\beq\label{eqtnm}
T_n(s_\lambda)
=\det (h_{k_i-i+j})_{1\leq i,j\leq \ell(\lambda)+1},
\eeq
where $k_1=n$, and $k_i=\lambda_{i-1}$ $(2\le i \le \ell(\lambda)+1)$.
\end{lem}
\noindent The proof of this lemma is elementary and we omit its details.

\begin{lem}\label{lemtres}
The transformation $T_0$ defined by~\eqref{eq-ts} satisfies
\beq\label{eq-tf}
T_0 (f)
\,=\, -\res_{z=\infty}f\bigl({\bf q}-[z^{-1}]\bigr) e^{\sum_{n\geq1} q_n z^n}\frac{dz}{z},
\quad \forall f\in\CC[[{\bf q}]].
\eeq
\end{lem}
\begin{prf} 
Due to the linearity of~$T_0$ and the fact that 
Schur polynomials form a basis of $\CC[[{\bf q}]]$, it suffices to show the validity of~\eqref{eq-tf} for all 
$s_\lambda$, $\lambda\in\YY$. 
On one hand, using Lemma~\ref{lemtnm}, we find
\beq
T_0(s_\lambda)
=\det (h_{k_i-i+j})_{1\leq i,j\leq \ell(\lambda)+1},
\eeq
where $k_1=0$, $k_i=\lambda_{i-1}$ ($2\leq i \leq \ell(\lambda)+1$).
On the other hand, observe, using~\eqref{eq-h}, that 
\beq
\sum_{k\geq0} h_k({\bf q}-[z^{-1}]) y^k
\,=\, \exp\bigg(\sum_{k\geq1}\big(q_k-\frac{1}{k z^k}\big) y^k\bigg)
\,=\, \exp\bigg(\sum_{k\geq1}q_k y^k\bigg) \, \Bigl(1-\frac y z\Bigr).
\eeq
This implies that
\beq
h_k({\bf q}-[z^{-1}])
\,=\, h_k({\bf q})-\frac1z h_{k-1}({\bf q}),\quad \forall \, k\geq1.
\eeq
Therefore, 
\begin{align}
s_\lambda({\bf q}-[z^{-1}]) 
= \det \Bigl(h_{\lambda_i-i+j}-\frac1z h_{\lambda_i-i+j-1}\Bigr)_{1\leq i,j\leq \ell(\lambda)}
= \sum_{k=0}^{\ell(\lambda)} C^{[k]} z^{-k},
\label{slambdaM}
\end{align}
where $C^{[k]}$ denotes the $(1,k+1)$ cofactor of 
the matrix $(h_{k_i-i+j})_{1\leq i,j\leq \ell(\lambda)+1}$.
Hence, 
\begin{align}
&-\res_{z=\infty}s_\lambda\bigl({\bf q}-[z^{-1}]\bigr) e^{\sum_{n\geq1} q_n z^n}\frac{dz}{z} \nn\\ 
&= -\res_{z=\infty}\Biggl(\sum_{k=0}^{\ell(\lambda)} C^{[k]}  z^{-k}\Biggr) \, \biggl(\sum_{k\geq0} h_k z^k \biggr)\frac{dz}{z} \nn\\
&= \sum_{k=0}^{\ell(\lambda)} h_k \, C^{[k]} 
\,=\, \det \left(h_{k_i-i+j}\right)_{1\leq i,j\leq \ell(\lambda)+1}. \nn
\end{align}
The lemma is proved.
\end{prf}

We notice that similarly with~\eqref{eq-tf} and~\eqref{eq-ts} the following formula is true:
\begin{align}
&-\res_{z=\infty} \, 
s_{(m_1,\dots,m_r|n_1,\dots,n_r)}
\bigl({\bf q}+[z^{-1}]\bigr) \, 
e^{-\sum_{n\geq1} q_n z^n} \, \frac{dz}{z} \nn \\
& =\left\{
\begin{array}{cc}
(-1)^r s_{(m_1+1,\dots,m_r+1|n_1-1,\dots,n_r-1)}(\bf q), \quad & n_r\geq1,\\
0, \quad &n_r=0.
\end{array}
\right.
\end{align}

\medskip
 
Now we are ready to prove Proposition~\ref{prp-qtq}. 

\medskip

\begin{prfn}{Proposition~\ref{prp-qtq}}
From Lemma~\ref{lem-gz} we know that there exists a series $g(z)\in\CC[[z^{-1}]]$ such that
\beq\label{eqtau12}
\tau_1 \tau_2 \,=\, -\res_{z=\infty} \, g(z) \, \tau_1\bigl({\bf q}-[z^{-1}]\bigr) \, e^{\sum_{n\geq1}q_nz^n} \, \frac{dz}z.
\eeq
Consider the expansion
\beq
\log g(z)=:b_0+\sum_{k\geq1} b_k z^{-k}.
\eeq
and denote $Q:=e^{\sum_{k\geq1} k b_k q_k}$. 
Using the condition that $\tau_2({\bf 0})$ equals 1,
we have $b_0=0$.
Then by using identity~\eqref{eqtau12} 
we obtain
\begin{align*}
\tau_1 \, \tau_2
=& -\res_{z=\infty} e^{\sum_{k\geq1}b_k z^{-k}} \tau_1({\bf q}-[z^{-1}]) \, e^{\sum_{n\geq1} q_n z^n} \frac{dz}z \\
=&-\res_{z=\infty}
\frac{e^{-\sum_{k\geq1}kb_k(q_k-\frac1{kz^k})}}{e^{-\sum_{k\geq1}k b_k q_k}} \tau_1({\bf q}-[z^{-1}]) \, 
e^{\sum_{n\geq1}q_nz^n}\frac{dz}z \\
=& - Q \, \res_{z=\infty} \bigl(Q^{-1}\tau_1\bigr)({\bf q}-[z^{-1}]) \, 
e^{\sum_{n\geq1}q_nz^n} \frac{dz}z,
\end{align*}
which, due to Lemma~\ref{lemtres}, gives
\[
Q^{-1}\tau_1\tau_2=T_0\left(Q^{-1}\tau_1\right).
\]
The proposition is proved.
\end{prfn}

\medskip 

\begin{lem}\label{lemqtq}
Let $Q=e^{\sum_{k\geq1}k b_k q_k}$ denote the multiplication operator. The following formula holds:
\beq
Q\circ T_0\circ Q^{-1}(s_\lambda)
=\sum_{n\geq0} \, h_n({\bf b}) \, T_n(s_\lambda).
\eeq
\end{lem}
\begin{prf}
Using Lemma~\ref{lemtres}, we have 
\begin{align}
Q\circ T_0\circ Q^{-1}\left(s_\lambda\right)
=&-\res_{z=\infty} e^{\sum_{k\geq1} b_k z^{-k}}
s_\lambda\bigl({\bf q}-[z^{-1}]\bigr) \,
e^{\sum_{n\geq1}q_n z^n}
\frac{dz}z \nn.
\end{align}
Then by substituting~\eqref{slambdaM} into this equality we obtain that 
\begin{align}
Q\circ T_0\circ Q^{-1}\left(s_\lambda\right) =&
-\res_{z=\infty}
\sum_{n,m\geq0} h_n({\bf b}) h_m({\bf q}) z^{m-n}
\sum_{k=0}^{\ell(\lambda)} C^{[k]} z^{-k} 
\frac{dz} z \nn\\
=&\sum_{n\geq 0} h_n({\bf b}) \sum_{k=0}^{\ell(\lambda)}h_{k+n}({\bf q}) \, C^{[k]}.
\end{align}
Here we recall that $C^{[k]}$ is the $(1,k+1)$ cofactor of 
the matrix $(h_{k_i-i+j})_{1\leq i,j\leq \ell(\lambda)+1}$ with $k_1=0$, $k_i=\lambda_{i-1}$ ($2\leq i \leq \ell(\lambda)+1$).
Obviously, for any $n\geq1$, $C^{[k]}$ is {\it also} the $(1,k+1)$ cofactor of 
the matrix $(h_{k_{n,i}-i+j})_{1\leq i,j\leq \ell(\lambda)+1}$
with $k_{n,0}=n$ and $k_{n,i}=\lambda_{i-1}$ ($2\leq i \leq \ell(\lambda)+1$).
We therefore conclude that 
\[
Q\circ T_0\circ Q^{-1}(s_\lambda) \, = \, \sum_{n\geq 0} h_n({\bf b}) \det (h_{k_{n,i}-i+j})_{1\leq i,j\leq \ell(\lambda)+1}.
\]
Combined with Lemma~\ref{lemtnm} the lemma is then proved.
\end{prf}

\medskip

We are now ready to prove Proposition~\ref{prp-sch-e}. 

\medskip

\begin{prfn}{Proposition~\ref{prp-sch-e}}
By using Proposition~\ref{prp-qtq} and Lemma~\ref{lemqtq}, 
it is easy to see that
\begin{align}
\tau_{{}_{\rm E}}
=\sum_{\lambda \in \YY} \pi_\lambda \, Q \circ T_0 \circ Q^{-1}(s_\lambda) 
=\sum_{\lambda \in \YY} \pi_\lambda \sum_{n\geq0} h_n({\bf b}) T_n(s_\lambda),
\label{tauea-2-section3}
\end{align}
which is nothing but an equivalent form of~\eqref{tauea}. 
Next we are to show~\eqref{tauea-1}. 
It is known~\cite{M} that 
the following identity holds true:
\beq\label{eqebq}
e^{\sum_{k\geq1}k q_k q'_k}
=\sum_{\lambda \in \YY} s_\lambda({\bf q}) s_\lambda({\bf q}').
\eeq
Then we have
\begin{align}
Q\circ T_0 \circ Q^{-1}(s_\lambda) 
=&\sum_{\nu\in\YY}s_\nu({\bf b})s_\nu({\bf q}) \, T_0
\Biggl(\sum_{\mu\in\YY}c_{\lambda \mu}^\beta s_\mu(-{\bf b}) s_\beta({\bf q})\Biggr)
\nn\\
=&\sum_{\substack{\alpha,\mu,\nu\in\YY \\ \beta\in \YY\backslash{\rm Ker}(T_0)}}
c_{\lambda\mu}^\beta c_{\nu,T_0(\beta)}^{\alpha}
 s_\mu(-{\bf b}) s_\nu({\bf b})s_{\alpha}({\bf q}). \nn
\end{align}
Formula~\eqref{tauea-1} is then obtained in a similar way with~\eqref{tauea-2-section3}. 
The proposition is proved.
\end{prfn}

For the reader's convenience we provide in below a few Littlewood--Richardson coefficients $c_{\alpha \mu}^\beta$ that appear in the above proof:
\begin{align*}
&s_{(3)}s_{(3)}=s_{(6)}+s_{(5,1)}+s_{(4,2)}+s_{(3,3)},\\
&s_{(3)}s_{(2,1)}=s_{(5,1)}+s_{(4,2)}+s_{(4,1,1)}+s_{(3,2,1)},\\
&s_{(3)}s_{(1,1,1)}=s_{(4,1,1)}+s_{(3,1,1,1)},\\
&s_{(2,1)}s_{(2,1)}=s_{(4,2)}+s_{(4,1,1)}+s_{(3,3)}+2s_{(3,2,1)}+s_{(3,1,1,1)}+s_{(2,2,2)}+s_{(2,2,1,1)},\\
&s_{(2,1)}s_{(1,1,1)}=s_{(3,2,1)}+s_{(3,1,1,1)}+s_{(2,2,1,1)}+s_{(2,1,1,1,1)},\\
&s_{(1,1,1)}s_{(1,1,1)}=s_{(2,2,2)}+s_{(2,2,1,1)}+s_{(2,1,1,1,1)}+s_{(1,1,1,1,1,1)}.
\end{align*}

For the case of the topological solution to the Burgers--KdV hierarchy, 
it was conjectured by Witten~\cite{W} and proved by Kontsevich the following identity~\cite{K1}:
\beq
\tau_1^{{}_{\rm WK}}
=\exp\left(\sum_{n\geq1}\sum_{i_1,\dots,i_n\geq0}\frac1{n!}
\prod_{k=1}^n(2i_k+1)!! \, q_{2i_k+1} \int_{\overline{\mathcal M}_{g,n}}\psi_1^{i_1}\cdots\psi_n^{i_n}\right),
\eeq
where $\tau_1^{{}_{\rm WK}}$ is defined by~\eqref{topo}--\eqref{string}. 
The second component $\tau_2^{{}_{\rm WK}}$ in the tau-tuple of the topological solution 
plays the role of the partition function for the open intersection numbers~\cite{B1,B2, PST}. 
From Corollary~\ref{corwk} and a straightforward computation, 
we obtain the first few terms of~$\tau_{{}_{\rm E}}^{{}_{\rm WK}}$ as follows:
\begin{align}
&\tau_{{}_{\rm E}}^{{}_{\rm WK}}
=1+\frac{41}{24} s_{(3)}-\frac{5}{24}s_{(2,1)}-\frac{7}{24} s_{(1,1,1)}+\frac{9241}{1152}s_{(6)}-\frac{385}{1152}s_{(5,1)}-\frac{455}{1152} s_{(4,1,1)} +\frac{205}{576}s_{(3,3)}\nn\\
&\quad+\frac{287}{576} s_{(3,2,1)}
+\frac{25}{1152} s_{(3,1,1,1)}-\frac{35}{576} s_{(2,2,2)}-\frac{385}{1152} s_{(2,1,1,1,1)}-\frac{455}{1152} s_{(1,1,1,1,1,1)}+\cdots,
\end{align}
which agree with the computations in~\cite{K}. 
We hope that Corollary~\ref{corwk} could be helpful in understanding open and closed intersection numbers.

\section{The extended generalized BGW tau-function}\label{sec4}

In this section, following~\cite{B2} and~\cite{BeY} 
we prove Theorem~\ref{thm-main}.
We also prove that the extended generalized BGW tau-function with some particularly chosen value of~$C$ is a polynomial.

Let $(\wc,\vc)$ and $(\taucone,\tauctwo)$ 
be defined in Introduction (cf.~\eqref{eq-ini}--\eqref{conditiontau20}). 
According to~\cite{BDY}, the series $\taucone$ is 
also the Sato type tau-function of the 
solution $(\wc,\vc)$ for the KdV hierarchy. Therefore,  
as mentioned in Section~\ref{sec2},
the $\psi_{{}_{\Theta(C)}}:=\psi_{{}_{\Theta(C)}}({\bf q};z)$ and $\psi^*_{{}_{\Theta(C)}}:=\psi^*_{{}_{\Theta(C)}}({\bf q};z)$ 
defined by 
\begin{align}
&\psi_{{}_{\Theta(C)}}=\frac{\taucone\bigl({\bf q}-[z^{-1}]\bigr)}{\taucone({\bf q})} \, e^{\sum_{n\geq1}q_n z^n},\quad 
\psi^*_{{}_{\Theta(C)}}=\frac{\taucone\bigl({\bf q}+[z^{-1}]\bigr)}{\taucone({\bf q})} \, e^{-\sum_{n\geq1}q_n z^n},
\end{align}
form a pair of wave and dual wave functions of~$\wc$ for the KdV hierarchy. To proceed we introduce the notations:
\begin{align}
&K_{{}_{\Theta(C)}}:=\taucone \, \psi_{{}_{\Theta(C)}},
\quad
f_{{}_{\Theta(C)}}(x,z):=\psi_{{}_{\Theta(C)}}\big|_{q_2=q_3=\cdots=0}, \label{eq-k}\\
&K^*_{{}_{\Theta(C)}}:=\taucone \, \psi^*_{{}_{\Theta(C)}},\quad
f^*_{{}_{\Theta(C)}}(x,z):=\psi^*_{{}_{\Theta(C)}} \big|_{q_2=q_3=\cdots=0}, \\
& L_0^{{}_{\rm Ext}}:=\sum_{n\geq1} n \, \tilde q_n \, \frac{\p}{\p q_n}+C,\quad \tilde q_n:=q_n-\delta_{n,1}.
\end{align}
\begin{lem}\label{lem-lk}
The following formulae hold true:
\begin{align}
&L_0^{{}_{\rm Ext}} K_{{}_{\Theta(C)}}
=(z\p_z-z)K_{{}_{\Theta(C)}},\quad \label{eq-lk1}\\
&L_0^{{}_{\rm Ext}} K^*_{{}_{\Theta(C)}}
=(z\p_z+ z)K^*_{{}_{\Theta(C)}}.\label{eq-lk2}
\end{align}
\end{lem}

\begin{prf}
Performing a shift ${\bf q}\mapsto{\bf q}-[z^{-1}]$ to identity~\eqref{eqstr1}, 
we obtain
\beq
\sum_{k\geq0} \left((2k+1) \, \tilde  q_{2k+1}-\frac{1}{z^{2k+1}}\right)\frac{\p\taucone}{\p q_{2k+1}}({\bf q}-[z^{-1}])
+C \, \taucone({\bf q}-[z^{-1}])=0.
\eeq
Then the left-hand side of~\eqref{eq-lk1} is equal to
\begin{align}
&e^{\sum_{n\geq1} q_n z^n}
\Biggl(\sum_{k\geq0}(2k+1) \, \tilde q_{2k+1}
\frac{\p \taucone({\bf q}-[z^{-1}])}{\p q_{2k+1}} 
 +\taucone({\bf q}-[z^{-1}]) \, 
\bigg(\sum_{n\geq1} n \tilde q_n z^n+C\bigg)\Biggr) \nn\\
&\quad =e^{\sum_{n\geq1}q_n z^n}\sum_{k\geq0}z^{-2k-1}
\frac{\p \taucone ({\bf q}-[z^{-1}])}{\p q_{2k+1}}
+K_{{}_{\Theta(C)}} \sum_{n\geq1} n \tilde q_n z^n \nn\\
&\quad =\left(z \p_z-z\right) K_{{}_{\Theta(C)}}. \label{lk3}
\end{align}
The proof for~\eqref{eq-lk2} is similar. 
The lemma is proved.
\end{prf}

The next lemma gives the explicit expressions of~$f_{{}_{\Theta(C)}}$ and~$f^*_{{}_{\Theta(C)}}$.
\begin{lem}\label{lem-formf}
The following formulae are true:
\begin{align}
f_{{}_{\Theta(C)}}
=e^{x z}\sum_{n\geq0} \frac{\theta_n}{z^n (1-x)^n},
\quad
f^*_{{}_{\Theta(C)}}
=e^{-x z}\sum_{n\geq0} \frac{(-1)^n\theta_n}{z^n (1-x)^n},\label{eq-f}
\end{align}
where $\theta_n:=(-1)^n\prod_{i=0}^{n-1}(C+\frac{i(i+1)}{2})/n!$, $n\geq0$.
\end{lem}
\begin{prf}
Setting $q_2=q_3=\cdots=0$ in~\eqref{eq-lk1}  we have
\beq\label{eq-fsz}
(x-1)\p_x f_{{}_{\Theta(C)}}=(z\p_z- z) f_{{}_{\Theta(C)}}.
\eeq
It follows that there exists a formal series $\theta(z)\in\CC[[z^{-1}]]$
such that
$f_{{}_{\Theta(C)}}=e^{x z} \theta\bigl(z\,(1-x)\bigr)$.
Setting $q_2=q_3=\cdots=0$ in formula~\eqref{eq-lpsi} we have
\beq\label{eq-pf}
\p_x^2 f_{{}_{\Theta(C)}}+\left(\frac{2C}{(1-x)^2} -z^2\right) f_{{}_{\Theta(C)}}=0.
\eeq
Therefore, we find that $\theta(z)$ satisfies the ODE:
\beq
\left(z^2 \p_z^2-2 z^2\p_z +2C\right) \theta(z)=0.
\eeq 
Solving this ODE with $\theta(z)\sim1$ as $z\to\infty$ we get the first equality. 
The proof of the second equality is similar. 
\end{prf}

\medskip

We are ready to prove Theorem~\ref{thm-main} using the method of~\cite{B2} improved in~\cite{BeY}. 

\medskip

\begin{prfn}{Theorem~\ref{thm-main}} Due to Lemma~\ref{lem-gz}, 
there exists a series $g(z)$ of the form~\eqref{gzform} satisfying
\beq
\tauctwo=-\res_{z=\infty}g(z)\frac{\taucone({\bf q}-[z^{-1}])}{\taucone({\bf q})}
e^{\sum_{n\geq1}q_n z^n}\frac{dz}{z}.
\eeq
Using~\eqref{conditiontau20} we find $g_0=1$.
Using Lemma~\ref{lem-lk} we have
\begin{align}
L_0^{{}_{\rm Ext}}(\tauce)
=\taucone \, \res_{z=\infty}  \bigl(g'(z)+g(z)\bigr) \, \psi_{{}_{\Theta(C)}} \, dz. \label{main1}
\end{align}
It follows from~\eqref{definitiontau12}, \eqref{eq-ini} and~\eqref{eqstr1} that
\beq
\left.\left(L_0^{{}_{\rm Ext}}(\tauce)\right)\right|_{q_2=q_3=\cdots=0}\equiv0.
\eeq
Thus  
taking $q_2=q_3=\cdots=0$ on the both sides of~\eqref{main1} we find 
\beq\label{a1}
\res_{z=\infty}\left(g'(z)+g(z)\right) f_{{}_{\Theta(C)}}(x,z) dz \equiv 0.
\eeq	
Since $\psi_{{}_{\Theta(C)}}$ satisfies equations~\eqref{eq-psi}, 
it follows from~\eqref{eq-szg} that
\beq
\res_{z=\infty} (g'(z)+g(z)) \psi_{{}_{\Theta(C)}}({\bf q};z) dz\equiv0.
\eeq
Following~\cite{BeY} we employ the bilinear identity~\eqref{eq-bili} with 
${\bf q}=(x,{\bf 0})$, ${\bf q}'=(x',{\bf 0})$, and we find
\beq\label{eq-szg}
\res_{z=\infty}  f^*_{{}_{\Theta(C)}}(x',z) \, f_{{}_{\Theta(C)}}(x,z) \, dz \, \equiv \, 0.
\eeq
Further taking $x'=0$ and using Lemma~\ref{lem-formf}
we find that 
\beq\label{a2}
\res_{z=\infty} \theta(-z) \, f_{{}_{\Theta(C)}}(x,z) \, dz \, \equiv \, 0. 
\eeq
Comparing~\eqref{a1} and~\eqref{a2}, and using the uniqueness nature we find that 
$g(z)$ must satisfy
\beq\label{linearg}
g'(z)+g(z) = \theta(-z).
\eeq
Obviously, solution to~\eqref{linearg} of the form~\eqref{gzform} is unique. 
By solving~\eqref{linearg} we finally obtain that 
$g(z)=1+\sum_{n\geq1}g_n^{{}_{\Theta(C)}}z^{-n}$ with 
$g_n^{{}_{\Theta(C)}}$ being given by~\eqref{eq-defgn}.
The identities~\eqref{eq-bgw12} and~\eqref{eqstr2} are proved. 
\end{prfn}

\medskip

It was shown in~\cite{A3} that 
if $C=-m(m+1)/2$ for some non-negative integer~$m$, 
the generalized BGW tau-function $\taucone$ is a polynomial. 
We find in the next proposition that when $m$ is even, 
the tau-function $\tauce$ is also a polynomial.

\begin{prp}\label{prp-poly} 
For $C=-m(m+1)/2$, $m\in \mathbb{Z}_{\geq 0}$, if $m$ is even, 
then $\tauce$ is a polynomial.
\end{prp}
\begin{prf}
Define $g(z)=\sum_{k\geq0} g_k^{{}_{\Theta(C)}} z^{-k}$.
Using Theorem~\ref{thm-main} we find
\beq\label{eq-gtri}
g(z)=\sum_{k=0}^{m-1}\frac{g_k^{{}_{\Theta(C)}}}{z^k}+a_m\sum_{k\geq m}\frac{(k-1)!}{z^k},
\eeq
where 
\beq
a_m=\sum_{k=1}^m \frac{\prod_{i=0}^{k-1}\Bigl(-\frac{m(m+1)}2+\frac{i(i+1)}2\Bigr)}{k!(k-1)!}.
\eeq
One can show that the number~$a_m$ satisfies
\beq
a_m=\frac1{m!}f^{(m)}(0),\quad f(z):=-z\left(1+z^2\right)^{-\frac32}.
\eeq
Since $m$ is even, we find $a_m$ is zero. 
Then from~\eqref{eq-gtri} it follows that $g(z)$ is a polynomial in~$z^{-1}$. 
Combining this with the polynomiality property of the component $\taucone$, 
one can see
\beq\label{onecansee}
g(z) \, \taucone\bigl({\bf q}-[z^{-1}]\bigr)\in\CC[{\bf q}][z^{-1}].
\eeq
By using Lemma~\ref{lem-gz} we obtain that
\beq\label{rewriteit}
\tauce \, = \, -\res_{z=\infty} \, 
g(z) \, \taucone\bigl({\bf q}-\left[z^{-1}\right]\bigr) \, e^{\sum_{n\geq1}q_n z^n} \, \frac{dz}{z}.
\eeq
We conclude from~\eqref{onecansee} and~\eqref{rewriteit} that $\tauce$ is a polynomial.
\end{prf}

\begin{appendices}

\section{The first few coefficients of $\tauce$}\label{app-a}

\begin{center}
\begin{spacing}{1.2}
\begin{tabular}{cc}
\toprule
$\lambda$ &  Coefficient of $s_\lambda$\\
\midrule
$(1)$ & $C$\\
$(2)$ & $ C (C+3)/2$\\
$(1,1)$ & $ C(C-1)/2$\\
$(3)$ & $ C(C+3)(C+7)/6$\\
$(2,1)$ & $ C(C+3)(2C-1)/6$\\
$(1,1,1)$ & $C(C+1)(C-3)/6$\\
$(4)$ & $C(C+3)(C+9)(C+10)/24$\\
$(3,1)$ & $C(C+3)(C^2+7C-2)/8$\\
$(2,2)$ & $C^2(C+1)(C+3)/12$\\
$(2,1,1)$ & $C(C+1)(C+3)(C-2)/8$\\
$(1,1,1,1)$ & $C(C+1)(C+3)(C-6)/24$\\
$(5)$ & $C(C+3)(C+10)(C^2+27C+186)/120$\\
$(4,1)$ & $C(C+3)(C+10)(2C^2-3C-30)/60$\\
$(3,2)$ & $C^2(C+1)(C+3)(C+8)/24$\\
$(3,1,1)$ & $C(C+1)(C+3)(C^2+6C-10)/20$\\
$(2,2,1)$ & $C^3(C+1)(C+3)/24$\\
$(2,1,1,1)$ & $C(C+1)(C+3)(2C^2-3C-30)/60$\\
$(1,1,1,1,1)$ & $C(C+1)(C+3)(C+6)(C-10)/120$\\
$(6)$ & $C(C+3)(C+10)(C+21)(C^2+31C+270)/720$\\
$(5,1)$ & $C(C+3)(C+10)(C^3+28C^2+201C-18)/144$\\
$(4,2)$ & $C^2(C+1)(C+3)(C+10)(C+11)/80$\\
$(4,1,1)$ & $C(C+1)(C+3)(C+10)(C^2+9C-9)/72$\\
$(3,3)$ & $C^2(C+1)(C+3)^2C(C+10)/144$\\
$(3,2,1)$ & $C^2(C+1)(C+3)(4C^2+34C+15)/180$\\
$(3,1,1,1)$ & $C(C+1)(C+3)(C^3+7C^2-15C-90)/72$\\
$(2,2,2)$ & $C^2(C+1)(C+3)(C^2+C+6)/144$\\
$(2,2,1,1)$ & $C^2(C+1)(C+3)(C^2+C-10)/144$\\
$(2,1,1,1,1)$ & $C(C+1)(C+3)(C+6)(C^2-5C-30)/144$\\
$(1,1,1,1,1,1)$ & $C(C+1)(C+3)(C+6)(C+10)(C-15)/720$\\
\bottomrule
\end{tabular}
\end{spacing}
\end{center}

\section{A novel proof of Alexandrov's theorem}\label{app}
In this section, we give a new proof of a result of Alexandrov~\cite{A2}.

Let $\CC((z^{-1}))$ be the linear space of formal series with finitely many terms of positive powers, 
and consider the Sato Grassmannian $GM$ as defined originally in~\cite{S}. 
A point $W\in GM$ is a subspace of $\CC((z^{-1}))$ which can be written as a linear span of a set of basis vectors. 
A particular useful choice of basis of $W$ is of the following form:
\beq\label{def-aff}
W={\rm Span}_{\CC} \Biggl\{z^l+\sum_{k\geq0}A_{k,l}z^{-k-1} \Biggr\}_{l\geq0}.
\eeq
Here the collection $\left(A_{k,l}\right)_{k,l\geq0}$ is called the \emph{affine coordinate} of $W$. 
Note that such a choice is not always possible: 
existence of basis of form~\eqref{def-aff} characterizes the \emph{big cell} (See~\cite{SW} for the details). Given a point $W\in GM$, the affine coordinate, if exist, must be unique.

Define a family of numbers as follows:
\beq
\pi_{(m_1,\dots,m_r|n_1,\dots,n_r)}(W):=(-1)^{\sum_{i=1}^r n_i}  \det(A_{m_i,n_j})_{1\leq i,j\leq r}.
\eeq
Then according to~\cite{BY,EH}, the formal series defined by
\beq\label{eq-tau-w}
\tau_{{}_{\rm W}}:=\sum_{\lambda\in\mathbb Y} \pi_\lambda(W) \, s_\lambda
\eeq
is a Sato type tau-function of the KP hierarchy, 
i.e., it satisfies the following bilinear identity:
\beq
\res_{z=\infty}\tau_{{}_{\rm W}}\bigl({\bf q}-[z^{-1}]\bigr) \, 
\tau_{{}_{\rm W}}\bigl({\bf q}'+[z^{-1}]\bigr) \, e^{\sum_{n\geq1}(q_n-q_n')z^n}dz=0.
\eeq

\begin{thm}[\cite{A2}]\label{thm-kp-tau}
Let $f(z)=1+\sum_{n\geq1}f_i \, z^{-n}$ be an arbitrary formal series in~$\CC[[z^{-1}]]$.
If $\tau$ is a Sato type tau-function of the KP hierarchy, so is
\beq\label{eq-tau-f}
\widetilde \tau:=-\res_{z=\infty} f(z)  \, \tau\bigl({\bf q}-[z^{-1}]\bigr) \, e^{\sum_{n\geq1}q_nz^n}\frac{dz}{z}.
\eeq
\end{thm}
\begin{prf}
By using Lemma~\ref{lemtres}, 
we can rewrite~\eqref{eq-tau-f} as
\beq\label{qtq-1}
\widetilde \tau=Q\circ T_0 \circ Q^{-1}(\tau),
\eeq
where $Q=e^{\sum_{k\geq1}k b_k q_k}$ denotes the multiplication operator 
with $b_1,b_2,\dots$ being numbers such that  
$\log f(z)=\sum_{k\geq1}b_k z^{-k}$.

Let $\tau$ be the Sato type tau-function given by~\eqref{eq-tau-w}. 
We first show $T_0(\tau)$ is a Sato type tau-function. 
Define a point in $GM$ as
\beq
\widetilde W:={\rm Span}_{\CC} \Biggl\{z^l-\sum_{k\geq0}A_{k+1,l-1}z^{-k-1} \Biggr\}_{l\geq1}
\eeq
It is obvious that the affine coordinates 
$\bigl(\widetilde A_{k,l}\bigr)_{k,l\geq0}$ of $\widetilde W$ are 
given by
\beq
\bigl(\widetilde A_{k,l}\bigr)_{k,l\geq0}
=\left(
\begin{array}{cccc}
\vdots & \vdots & \vdots & \vdots\\
0 & -A_{20} & -A_{21} & \cdots\\
0 & -A_{10} & -A_{11} & \cdots
\end{array}
\right).
\eeq
Recall the expression~\eqref{eq-ts} of $T_0$. 
Then by using~\eqref{eq-tau-w} and observing that 
\beq
\det \bigl(\widetilde A_{m_i,n_j} \bigr)_{1\leq i,j\leq r}=
\left\{
\begin{array}{cc}
(-1)^r \det \bigl(A_{m_i+1,n_j-1} \bigr)_{1\leq i,j\leq r},\quad & n_r\geq1,\\
0, \quad & n_r=0,
\end{array}
\right.
\eeq
one can obtain that
\begin{align*}
T_0(\tau)=\tau_{{}_{\widetilde{W}}}.
\end{align*}
Hence $T_0(\tau)$ is Sato type tau-function of the KP hierarchy. 
In other words, the transformation~$T_0$ has the property that it maps an arbitrary Sato type tau-function of the KP hierarchy to another one.
Obviously, the operator $Q$ also has this property. 
Combining with~\eqref{qtq-1}, 
we therefore conclude that the formal series $\widetilde\tau$ is also a Sato type tau-function. 
The theorem is proved.
\end{prf}

Let us give a direct consequence of Proposition~\ref{prp-sch-e} and Corollary~\ref{coraffine}. 
For an arbitrary KP tau-function in the big cell with affine coordinates $A_{m,n}$ ($m,n\geq0$) and for arbitrary
$g_k$ ($k\geq0$) with $g_0=1$,   
define $c_{m,n}$ by~\eqref{primea} and $\pi_\lambda$ by~\eqref{ZhouBaloghYang}. 
Then for any partition $\lambda=(m_1,\dots m_r|n_1,\dots,n_r)$, the following identities are true:
\beq\label{id1}
\det (c_{m_i,n_j})_{1\leq i,j\leq r} 
=\sum_{\substack{0\leq k \leq |\lambda|\\ \mu\in \mathcal T_k^{-1}(\lambda)}} g_k  \, \pi_\mu=\sum_{\substack{\alpha,\mu,\nu\in\YY \\ \beta\in\YY\backslash{{\rm Ker}}(T_0)}}
\pi_\alpha \, c_{\mu\alpha}^\beta \, c_{\nu,T_0(\beta)}^\lambda \, s_ \mu(-{\bf b}) \, s_\nu ({\bf b}),
\eeq
where ${\bf b}:=(b_1,b_2,\dots)$ are defined via $e^{\sum_{k\geq1} b_k z^{-k}}=\sum_{n\geq0} g_n z^{-n}$. 

\end{appendices}

\medskip

\noindent School of Mathematical Sciences, USTC, Hefei 230026, P.R. China

\smallskip

\noindent diyang@ustc.edu.cn, zhouch@ustc.edu.cn

\end{document}